\documentclass[twocolumn,showpacs,amsmath,amssymb]{revtex4}

\usepackage{graphicx} 
\usepackage{dcolumn}
\usepackage{bm}

\newcommand{\be }{\begin{equation}}
\newcommand{\bea }{\begin{eqnarray}}
\newcommand{\bh }{\begin{displaymath}}
\newcommand{\en }{\end{equation}}
\newcommand{\ena }{\end{eqnarray}}
\newcommand{\eh }{\end{displaymath}}

\begin{document}

\title{Electrostatic spherically symmetric configurations in gravitating nonlinear electrodynamics}

\author{J. Diaz-Alonso}\email{joaquin.diaz@obspm.fr}
\author{D. Rubiera-Garcia}\email{rubieradiego@uniovi.es}
\address{LUTH, Observatoire de Paris, CNRS, Universit\'e Paris
Diderot. 5 Place Jules Janssen, 92190 Meudon, France}
\address{Departamento de F\'isica, Universidad de Oviedo. Avda.
Calvo Sotelo 18, E-33007 Oviedo, Asturias, Spain}

\date{\today}

\begin{abstract}

We perform a study of the gravitating electrostatic spherically symmetric (G-ESS) solutions of Einstein field equations minimally coupled to generalized non-linear abelian gauge models in three space dimensions. These models are defined by lagrangian densities which are general functions of the gauge field invariants, restricted by some physical conditions of admissibility. They include the class of non-linear electrodynamics supporting ESS non-topological soliton solutions in absence of gravity. We establish that the qualitative structure of the G-ESS solutions of admissible models is fully characterized by the asymptotic and central-field behaviours of their ESS solutions in flat space (or, equivalently, by the behaviour of the lagrangian densities in vacuum and on the point of the boundary of their domain of definition, where the second gauge invariant vanishes). The structure of these G-ESS configurations for admissible models supporting divergent-energy ESS solutions in flat space is qualitatively the same as in the Reissner-Nordstr\"om case. In contrast, the G-ESS configurations of the models supporting finite-energy ESS solutions in flat space exhibit new qualitative features, which are discussed in terms of the ADM mass, the charge and the soliton energy. Most of the results concerning well known models, such as the electrodynamics of Maxwell, Born-Infeld and the Euler-Heisenberg effective lagrangian of QED, minimally coupled to gravitation, are shown to be corollaries of general statements of this analysis.

\end{abstract}


\maketitle

\section{Introduction}

For more than two decades a great deal of attention has been focused on the study of gravitating field configurations. This renewed interest is partially motivated by the discovery that generalized field theories, such as the Born-Infeld (BI) lagrangian (introduced in the 30's to remove the divergence of the electron self-energy in classical electrodynamics \cite{BI34}) and some non-abelian BI-like versions arise, together with the gravitational field, in the low-energy limit of string theory \cite{fradkin85}. But the interest in solutions of the Einstein equations coupled to different kinds of fields has other motivations. Indeed, in a four-dimensional flat space, several non-existence theorems \cite{deser76} restrict drastically the class of field theories supporting soliton solutions. However, through the coupling to gravity this obstruction can be removed and thus gravitating particle-like solutions can be found in theories that do not support soliton solutions in flat space, such as the pure Yang-Mills one \cite{bart88}. On the other hand, theories supporting soliton solutions in flat space, as the abelian and non-abelian BI models, have been extended to curved space leading to black hole-like solutions (see \cite{v-d-g} for a review). It is also worth mentioning the so-called soliton stars, gravitating coherent quantum field states with the features of non-topological solitons \cite{lee87}.

In the context of electromagnetic field theories, generalizations of the Reissner-Nordstr\"om (RN) solution of the Einstein-Maxwell field equations have been developed for some particular examples of non-linear electrodynamics (NED) coupled to gravity. The first finding in this field dates back to the work of Hoffmann \cite{hoffmann37}. In the eighties, aside from the electrically charged black hole solutions in the Einstein-BI system \cite{garcia84}, similar configurations were found in more general BI-type electrodynamics \cite{oliveira94}, minimally coupled to gravity. These studies were later extended by the inclusion of a dilaton field \cite{tamaki00}, the analysis of black holes in Anti-de Sitter spaces \cite{fernando03} and its generalization to higher-dimensional cases \cite{dey04}. Besides BI-type fields the coupling to gravity of other NED lagrangians has been also considered. Let us mention the gravitating electrostatic, monopole and dyon solutions of the Euler-Heisenberg (EH) effective lagrangian of QED \cite{yajima01}, the logarithmic lagrangian of Ref.\cite{soleng95} and the class of models obtained as powers of the Maxwell lagrangian, studied in Ref.\cite{hassaine07}.

In the search for regular (black hole or not) solutions, a theorem \cite{bronnikov80} forbids their existence for electrostatic fields in NED models having the Maxwell weak field limit. But this theorem can be circumvented for purely magnetic solutions \cite{bronnikov01}, or through a coupling between different fields \cite{burinskii}. Moreover, some gravitating NED models were reported to lead to electrically charged non-singular black hole solutions \cite{a-b}, but they are based on unphysical (multivalued) lagrangian density functions \cite{bronnikov0}.

In this work we perform a general study of G-ESS solutions for a large class of NED models minimally coupled to gravity. The lagrangian densities for these abelian gauge fields are defined as arbitrary functions of the two standard gauge invariants and include most of the aforementioned models. This class is restricted by some physically reasonable ``admissibility" conditions, such as the regularity and uniqueness of the lagrangian densities, the positivity of the energy, the parity invariance and the asymptotic vanishing of the ESS fields (faster than $1/r$). These admissible models can be exhaustively classified into two sets. The first one includes the models with ESS solutions in flat space which are energy-divergent, owing to their central-field behaviours (the Maxwell theory being the simplest example). The second one includes the models supporting finite-energy ESS solutions in flat space, which were extensively analyzed in Ref.\cite{dr08}. The structure of the G-ESS solutions of the first set is shown to be the same as that of the RN solutions of the Einstein-Maxwell field equations. The ``soliton-supporting" NED models, when coupled to gravity, lead to new qualitative features of the G-ESS solutions (single-horizon black holes, ``black points", etc) which we analyze and classify in terms of the charge, ADM mass and flat-space soliton energy. We shall show that the central-field behaviours of the corresponding ESS solutions in flat space which, as already mentioned, allowed the determination and classification of the full set of this kind of admissible models \cite{dr08}, allow also the complete characterization of the corresponding gravitational configurations. This central-field behaviour is determined by the specification of the form of the admissible lagrangian densities on the boundary of their domain of definition for vanishing values of the second gauge invariant. Consequently, the gravitational structure of these solutions is fully characterized by admissibility and this boundary behaviour, regardless of the explicit expressions of the lagrangian densities elsewhere. The G-ESS solutions for some models studied in the literature (as the gravitating BI \cite{garcia84} and EH \cite{yajima01}) are exposed as representative examples of the different classes introduced here, and their gravitational features are shown to be immediate consequences of the general results derived in the present analysis.

\section{Models and field equations}

Let us precise our conventions and conditions. In $(3+1)$ dimensions the two quadratic invariants of the abelian field, built from the Maxwell field strength tensor $F_{\mu\nu} = \partial_{\mu}A_{\nu} - \partial_{\nu}A_{\mu}$ and its dual $F_{\mu\nu}^* = \frac{1}{2}\epsilon_{\mu\nu\alpha\beta}F^{\alpha\beta}$, are
defined as $X = -\frac{1}{2}F_{\mu\nu}F^{\mu\nu} = \vec{E}^2-\vec{B}^2$ and $Y = -\frac{1}{2}F_{\mu\nu} F^{*\mu\nu} = 2\vec{E}\cdot\vec{B}$, where the electric and magnetic fields are $E^i=-F^{0i}$ and $B^i = -\frac{1}{2}\epsilon^{ijk}F_{jk}$, respectively. The dynamics of these fields is governed by lagrangian density functions $\varphi(X,Y)$ of these invariants. For physical admissibility \cite{dr08} these functions must be
\textit{single-branched, continuous and derivable} on their domains of definition ($\Omega \in \Re^{2}$) of the $X-Y$ plane, which are assumed to be open, connected and including the vacuum ($(X=0,Y=0) \in \Omega$). The regularity (for $r \neq 0$) of the ESS solutions requires also $\varphi(X,Y)$ to be of class $C^{1}$ on the line $(X>0, Y=0) \bigcap \Omega$ and $\partial \varphi/\partial X$ to be strictly positive there. We also require $\varphi(X,Y)$ to be symmetric in the second argument ($\varphi(X,Y) = \varphi(X,-Y)$), in order to implement parity invariance. Moreover, the requirement of the positive definite character of the energy is essential in the present context and leads to the minimal \textit{necessary and sufficient} condition on the energy density $\rho$, (obtained from the symmetric energy-momentum tensor of Eq.(\ref{eq:(2-4)}) below)

\be
\rho \geq \left(\sqrt{X^{2}+Y^{2}} + X\right) \frac{\partial \varphi}{\partial X}+ Y\frac{\partial \varphi}{\partial Y} - \varphi(X,Y) \geq 0,
\label{eq:(2-1)}
\en
to be satisfied in the entire domain of definition $\Omega$.

These models are minimally coupled to gravity through the action

\be
S=S_{G} + S_{NED} = \int d^4x \sqrt{-g}\left[\frac{R}{16\pi G} - \varphi(X,Y)\right],
\label{eq:(2-2)}
\en
where, as usual, $R$ is the scalar of curvature and $g$ is the determinant of the metric tensor $g_{\mu\nu}$. The associated Euler field equations, together with the Bianchi identities for the electromagnetic field, take the form

\bea
\nabla_{\mu}\left(\varphi_{x} F^{\mu\nu} + \varphi_{y} F^{*\mu\nu}\right) &=& 0 \nonumber\\
\nabla_{\mu}\ F^{*\mu\nu}&=&0,
\label{eq:(2-3)}
\ena
where $\varphi_{x} = \frac{\partial \varphi}{\partial X}$ and $\varphi_{y} = \frac{\partial \varphi}{\partial Y}$. On the other hand, the gauge-invariant symmetric energy-momentum tensor reads

\bea
T_{\mu\nu} &=& \frac{-2}{\sqrt{-g}}\frac{\delta S_{NED}}{\delta g^{\mu\nu}} = \nonumber \\
&=& 2\left(\varphi_{x} F_{\mu\alpha}F^{\alpha}_{\nu} - \varphi_{y} F_{\mu\alpha}F^{*\alpha}_{\nu}\right) - g_{\mu\nu} \varphi(X,Y).
\label{eq:(2-4)}
\ena
For ESS solutions we have

\be
\vec{E} = E(r)\frac{\vec{r}}{r} \hspace{5pt};\hspace{5pt} \vec{B} = 0.
\label{eq:(2-4)bis}
\en
In this case the energy-momentum tensor (\ref{eq:(2-4)}) satisfies $T_{0}^{0} = T_{1}^{1}$ and, as a direct consequence, the metric can be cast into the Schwarzschild-like form

\be
ds^{2} = \lambda(r) dt^{2} - \frac{dr^{2}}{\lambda(r)} - r^2 d\Omega^{2},
\label{eq:(2-5)}
\en
where $d\Omega^{2} = d\theta^{2} + \sin^{2} \theta d\phi^{2}$. In this coordinate system the non-vanishing components of the symmetric energy-momentum tensor for ESS fields read

\be
T^{0}_{0} = T^{1}_{1} = 2\varphi_{x} E^{2} - \varphi \hspace{5pt}; \hspace{5pt} T^{2}_{2} = T^{3}_{3} = -\varphi,
\label{eq:(2-6)}
\en
from which the Einstein equations are immediately written. Under the conditions of admissibility stated above (in particular, Eq.(\ref{eq:(2-1)})) it can be easily shown that the Weak Energy Condition $T_{\mu\nu} \xi^{\mu}\xi^{\nu} \geq 0$ (where $\xi^{\mu}$ is a time-like vector) automatically holds. This condition ensures that a time-like observer measures a non-negative energy density of the field at any point.

The generalized Maxwell equations (\ref{eq:(2-3)}) for these ESS fields take now the same form, in the Schwarzschild coordinate system, as the corresponding equations in the flat space problem (in spherical coordinates) and have the same first-integral, which reads

\be
r^{2} \varphi_{x} E(r) = q,
\label{eq:(2-7)}
\en
where $q$ is an integration constant, identified as the electric charge \cite{dr08} (without loss of generality, owing to the positivity of $\varphi_{x}$ on $Y=0$, which is a consequence of Eq.(\ref{eq:(2-1)}), we consider the case $q>0$, $E(r)>0$ only). The invariant $X$ is now given by $X = E^{2}$ (note that the second set of equations in (\ref{eq:(2-3)}) is identically satisfied for these fields). Let us emphasize the fact that the first-integral (\ref{eq:(2-7)}) \textit{does not depend explicitly on the metric function} $\lambda(r)$. It determines implicitly the form of the field $E(r)$, once the expression of the lagrangian function $\varphi(X,Y)$ is specified. For a given lagrangian this form is the same, in the Schwarzschild coordinate system, as in the absence of gravitation in spherical coordinates. Moreover, the expression of $T^{0}_{0}(r)$ (as well as the other components of the energy-momentum tensor) in (\ref{eq:(2-6)}) is also the same as in the flat space case. Consequently, the results of the analysis of Ref.\cite{dr08} on generalized NEDs can be immediately translated to this gravitating problem. Let us recall some main results of this reference which are pertinent for the present study. We have characterized the lagrangians of the NEDs by the behaviours of their ESS solutions at the origin and at infinity, in order to establish the conditions for the convergence of the integral of energy. Let us assume in both limits a behaviour of the form

\be
E(r) \sim \left(\frac{\beta}{q}\right)^{p/2} r^{p},
\label{eq:(2-7)bis}
\en
(with $p\neq 0$), $\beta$ being a characteristic constant for each model, given by the limits

\be
\beta = \lim \left(X^{\frac{p+2}{2p}}\frac{\partial \varphi}{\partial X}\right),
\label{eq:(2-7)e}
\en
calculated as $X \rightarrow 0$ or as $X \rightarrow \infty$ for the asymptotic or central-field behaviours, respectively. We imposed $p < -1$ as $r \rightarrow \infty$ (in order for the integral of energy to converge as $r \rightarrow \infty$) and we distinguished three cases, corresponding asymptotically to a slower than coulombian damping ($-2 < p < -1$, \underline{case B1}), a coulombian damping ($p = -2$, \underline{case B2}) and a faster than coulombian damping ($p < -2$, \underline{case B3}) of the field. As $r \rightarrow 0$ we considered the models with $-1 < p \leq 0$ (the only admissible ones compatible with the convergence of the integral of energy at the center) and we distinguished two cases. The \underline{case A1} ($-1 < p < 0$), which corresponds to ESS fields divergent (but integrable) at the center, behaving as in Eq.(\ref{eq:(2-7)bis}), and the \underline{case A2} ($p = 0$), which corresponds to a constant value of the field at the center, where it behaves as

\be
E(r \rightarrow 0) \sim a - b r^{\sigma},
\label{eq:(2-7)ter}
\en
$a$ and $\sigma$ being positive constants characteristic of the model (for example, for the BI model, $\sigma = 4$ and the maximum field strength is $E(0) = a$). The constant $b$ is related to the charge of the particular ESS solution through

\be
\alpha = \lim_{X\rightarrow a^{2}} \left[a(a-\sqrt{X})^{2/\sigma}\dfrac{\partial \varphi}{\partial X}\right] = q b^{2/\sigma},
\label{eq:(2-7)quart}
\en
$\alpha$ being a characteristic parameter of the model.

The behaviours of the lagrangian densities $\varphi(X,Y=0)$ corresponding to the asymptotic behaviours of the finite-energy ESS solutions take the generic form (see Fig.1)

\be
\varphi(X,Y=0) \sim \frac{\beta}{\gamma} X^{\gamma},
\label{eq:(2-7)f}
\en
around the vacuum ($X = E^{2}(r \rightarrow \infty) \sim r^{p} \rightarrow 0$), where $\gamma = \frac{p-2}{2p}$. For the integral of energy to converge asymptotically this parameter must range in the interval $1/2 < \gamma < 3/2$, with $1 < \gamma < 3/2$ in \underline{case B1}, $\gamma = 1$ in \underline{case B2} and $1/2 < \gamma < 1$ in \underline{case B3}.

The same expression (\ref{eq:(2-7)f}) with $\gamma > 3/2$ describes the behaviour of the lagrangian density for large $X$ in \underline{case A1}. This function diverges on the boundary of $\Omega$ at the point $Y=0$ and $X = E^{2}(r \rightarrow 0) \sim r^{p} \rightarrow \infty$, exhibiting there a vertical parabolic branch. In \underline{case A2} the boundary of $\Omega$ at $Y=0$ is the point ($X = E^{2}(r=0) = a^{2}, Y=0$) and the lagrangian density behaves there (for $\sigma \neq 2$) as

\be
\varphi(X) \sim \frac{2\alpha \sigma}{2-\sigma}(a - \sqrt{X})^{\frac{\sigma-2}{\sigma}} + \varphi(a^{2},Y=0),
\label{eq:(2-7)six}
\en
exhibiting a vertical asymptote in this point for $\sigma < 2$, or taking a finite value there with infinite slope for $\sigma > 2$. For $\sigma = 2$ the lagrangian density behaves as

\be \varphi(X) \sim -\alpha \ln(a - \sqrt{X}),
\label{eq:(2-7)sev}
\en
and exhibits a vertical asymptote at $X = a^{2}$.

In putting aside these conditions we are lead to six patterns of models for the class of admissible NEDs supporting (flat-space) finite-energy ESS solutions. Nevertheless, here we are also interested in the gravitating versions of NEDs supporting energy-divergent (due to the central field behaviour) ESS solutions in flat space, as the RN solution of the Einstein-Maxwell field equations. Consequently, in the sequel we shall consider also cases where the ESS field solutions diverge at the center faster than $1/r$ ($p < -1$ in Eq.(\ref{eq:(2-7)bis}) for $r \rightarrow 0$,corresponding to $1/2 < \gamma < 3/2$ in Eq.(\ref{eq:(2-7)f}) for large $X$).

\begin{figure}[h]
\includegraphics[width=8.6cm,height=6.5cm]{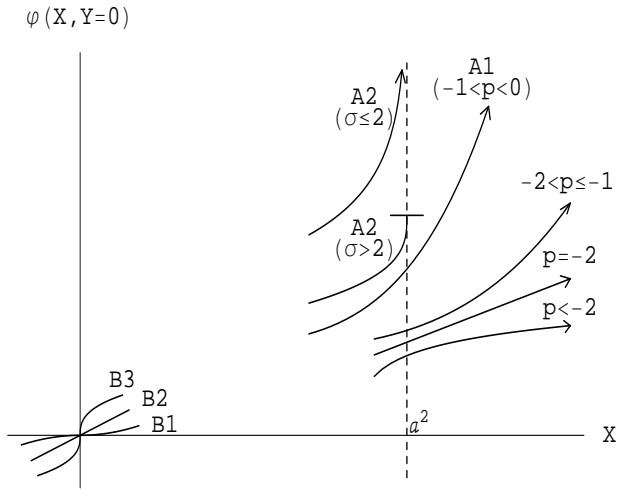}
\caption{\label{fig:epsart} Behaviours of the lagrangian densities of admissible NEDs for $Y = 0$. Around the vacuum ($X=0$) they determine the asymptotic behaviours of the ESS solutions, necessary for the integral of energy to converge at infinity (cases B1, B2 and B3). The central behaviour of the ESS solutions characterizes the structure of the corresponding G-ESS configurations. This behaviour is governed by the form of the lagrangian densities on the boundary of their domain of definition. The cases A1 and A2 correspond to ESS solutions for which the integral of energy converges at the center. In case A2 the boundary is reached at finite $X$ and the lagrangian densities exhibit there a vertical asymptote (if $\sigma \leq 2$) or an absolute maximum with divergent slope (if $\sigma > 2$). In case A1 the boundary is at infinity and the lagrangian densities diverge asymptotically as $\varphi(X \rightarrow \infty) \sim X^{\gamma}$ with $\gamma > 3/2$, exhibiting a vertical parabolic branch there. Lagrangian densities with similar asymptotic form and $1/2 < \gamma \leq 3/2$ ($p \leq -1$), leading to ESS fields with a divergent integral of energy (at the center), have also been drawn.}
\end{figure}

Note that the field equations can now be deduced from the Bianchi identity $\nabla_{\mu}G^{\mu\nu}=0$ and, consequently, (\ref{eq:(2-7)}) is also a first-integral for the Einstein equations. By combining Einstein equations and the first-integral we are lead to the system (we take $G=1$)

\bea
\frac{d}{dr}\left(r\lambda(r) - r\right) &=& -8\pi r^{2} T^{0}_{0} = -8\pi r^{2}\left(2\varphi_{x} E^{2} - \varphi\right)\nonumber \\
\frac{d^{2}}{dr^{2}}\left(r\lambda(r)\right) &=& -16\pi r T^{2}_{2} = 16\pi r \varphi,
\label{eq:(2-8)}
\ena
whose compatibility can be easily established. Thus the solution of the G-ESS problem reduces to solve the first of Eqs.(\ref{eq:(2-8)}), when $E(r)$ is determined from Eq.(\ref{eq:(2-7)}), once the form of the lagrangian function is specified. The integration of (\ref{eq:(2-8)}) leads to the expression

\be
\lambda(r,q) = 1 - \frac{2M}{r} + \frac{2\varepsilon_{ex}(r,q)}{r},
\label{eq:(2-9)}
\en
where $M = \frac{1}{2} \lim_{r\rightarrow\infty}[r(1 - \lambda(r))]$ is an integration constant, identified as the ADM mass, and

\be
\varepsilon_{ex}(r,q) = 4\pi \int_{r}^{\infty} R^{2} T^{0}_{0}(R,q) dR,
\label{eq:(2-10)}
\en
is the energy in flat space outside the sphere of radius $r$ (by convenience we shall call ``integrals of energy" the expressions similar to (\ref{eq:(2-10)})). The integrand of (\ref{eq:(2-10)}) in terms of the field $E(r,q)$ can be obtained from Eqs.(\ref{eq:(2-6)}) and (\ref{eq:(2-7)}) and reads

\bea \label{eq:(2-11)}
r^{2} T^{0}_{0}(r,q) &=& 2qE(r,q) - r^{2}\varphi = \\
&=& \nonumber 2q\left[E(r,q) + r^{2}\int_{r}^{\infty} \frac{dE(R,q)/dR}{R^{2}} dR\right],
\ena
whose asymptotic behaviour can be shown to be the same of $E(r,q)$. Then the integral defining $\varepsilon_{ex}(r,q)$ is convergent for $r > 0$. The solution (\ref{eq:(2-9)}) contains the charge $q$ and the mass $M$ as arbitrary constants. The dependence of the integral of energy on the charge is given by

\be
\varepsilon_{ex}(r,q) = q^{3/2} \varepsilon_{ex}\left(\frac{r}{\sqrt{q}}, q = 1\right).
\label{eq:(2-12)}
\en
Another important result is that, for any admissible model, $\varepsilon_{ex}(r,q)$ is a \textbf{monotonically decreasing and concave function of $r$}. This can be proven by double derivation of (\ref{eq:(2-10)}), taking into account (\ref{eq:(2-11)}) and the monotonically decreasing character of $E(r)$, which can be established from (\ref{eq:(2-7)}) and the admissibility conditions (see Ref.\cite{dr08}). Another useful expression for the exterior integral of energy in terms of the electrostatic potential can be obtained from Eqs.(\ref{eq:(2-10)}), (\ref{eq:(2-11)}) and (\ref{eq:(2-7)}) and reads

\be
\varepsilon_{ex}(r,q) = \dfrac{16\pi q}{3} A_{0}(r,q) - \dfrac{4\pi}{3}r^{3}T_{0}^{0}(r,q).
\label{eq:(2-12)bis}
\en

All these results allow the complete characterization of the structure of the G-ESS solutions of any physically meaningful gravitating NED of the form (\ref{eq:(2-2)}). We shall begin by the analysis of the family of those models leading to energy-divergent ESS solutions in flat space.

\section{Models with energy-divergent ESS solutions in flat space}

Let us return to Eq.(\ref{eq:(2-9)}) and look for solutions $r_{h}$ of $\lambda(r) = 0$, which give the radii of the horizons present in the configuration. In obtaining the solutions $r_{h} \neq 0$ we can solve the equivalent equation

\be
M - \frac{r}{2} = \varepsilon_{ex}(r,q),
\label{eq:(2-13)}
\en
which simplifies greatly the discussion and will give the mass-horizon-radius relation for the black holes. Indeed, the solutions of (\ref{eq:(2-13)}) are the intersection points between the curves $y = \varepsilon_{ex}(r,q)$ and the beam of straight lines $y = M - r/2$ in the $r-y$ plane (see Fig.2). The monotonic character and the concavity of $y = \varepsilon_{ex}(r,q)$ for any value of $q$ lead to three possibilities:

\textbf{(1)} There is a family of extreme black hole solutions (with parameter $q$) which correspond to the points of the exterior energy curves with slope $-1/2$. The corresponding radii of the horizons are the solutions $r_{hextr}(q)$ of the equations

\be
8\pi r^{2} T^{0}_{0}(r,q) = 1,
\label{eq:(2-14)}
\en
and the associated values of the ADM masses, $M_{extr}(q)$, are given by replacing $r_{hextr}(q)$ into the general mass-horizon radius relation (\ref{eq:(2-13)}). Using Eqs.(\ref{eq:(2-12)bis}) and (\ref{eq:(2-14)}) we are lead, in terms of the electrostatic potential, to the extreme black hole mass formula

\be
M_{extr}(q) = \dfrac{r_{hextr}(q)}{3} + \dfrac{16\pi q}{3} A_{0}(r_{hextr}(q),q),
\label{eq:(2-14)bis}
\en
which is valid for G-ESS fields vanishing asymptotically faster than $1/r$. As easily shown from Eq.(\ref{eq:(2-14)}), the scaling law (\ref{eq:(2-12)}) and the shape of $\varepsilon_{ex}(r)$, as $q$ increases from $0$ to $\infty$, $r_{hextr}(q)$ and $M_{extr}(q)$ get increased in the same range.

\begin{figure}[h]
\includegraphics[width=8.6cm,height=5.8cm]{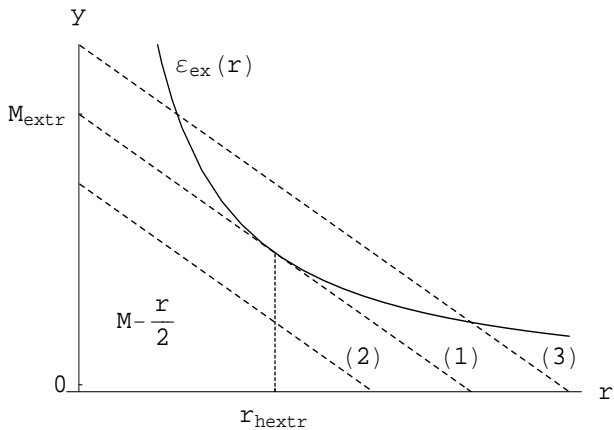}
\caption{\label{fig:epsart} Exterior integral of energy $\varepsilon_{ex}(r)$ for NEDs with (flat-space)
energy-divergent ESS solutions. The straight lines correspond to different values of the ADM mass $M$. The cut points give the horizons of the black hole solutions: (1) tangency point corresponding to single-horizon extreme black holes ($M=M_{extr}(q)$); (2) naked singularities ($M<M_{extr}(q)$); (3) black holes with two horizons ($M>M_{extr}(q))$.}
\end{figure}

\textbf{(2)} For configurations with values of $M < M_{extr}(q)$ there are no horizons and the solutions have a naked singularity at the center.

\textbf{(3)} For configurations with values of $M > M_{extr}(q)$ there are always two horizons (event and Cauchy) for each configuration. The corresponding radii are the solutions of (\ref{eq:(2-13)}). In both (2) and (3) cases, the limit $M \rightarrow M_{extr}(q)$ is singular (see Ref.\cite{carroll09}).

Let us stress that all these results on the gravitational properties of the ESS solutions at finite $r$ are independent of their asymptotic behaviours. They are similar to those of the RN solution, corresponding to the Maxwell lagrangian density

\be
\varphi(X,Y) = \alpha X,
\label{eq:(2-15)}
\en
($\alpha$ being a constant) which characterizes the pattern for the structure of the gravitational field of the solutions of this family. In this case Eqs.(\ref{eq:(2-13)}) and (\ref{eq:(2-14)}) can be solved explicitly and, aside from Eqs.(\ref{eq:(2-6)}) and (\ref{eq:(2-7)}), lead to the well known results (see, for example, Ref.\cite{ortin})

\bea
E(r,q) &=& \dfrac{q}{\alpha r^{2}}\hspace{.05cm}; \hspace{.05cm} r^{2} T^{0}_{0}(r,q) = \dfrac{q^{2}}{\alpha r^{2}}\hspace{.05cm}; \hspace{.05cm} \varepsilon_{ex}(r,q) = \dfrac{4\pi q^{2}}{\alpha r} \nonumber\\
r_{h}(q) &=& M \pm \sqrt{M^{2} - \frac{8\pi q^{2}}{\alpha}} \\ r_{hextr}(q) &=& M_{extr}(q) = q\sqrt{\frac{8\pi}{\alpha}}. \nonumber
\label{eq:(2-15)bis}
\ena

\section{Models with finite-energy ESS solutions in flat space}

In this case the integral of energy converges at the center ($r \rightarrow 0$) and we can define both the exterior and the interior integrals of energy for a given sphere

\bea
\varepsilon_{in}(r,q) &=& 4\pi \int_{0}^{r} R^{2} T^{0}_{0}(R,q) dR \nonumber \\
\varepsilon_{ex}(r,q) &=& 4\pi \int_{r}^{\infty} R^{2} T^{0}_{0}(R,q) dR,
\label{eq:(2-17)}
\ena
which are related through

\be
\varepsilon(q) = \varepsilon_{in}(\infty,q) = \varepsilon_{ex}(0,q) = \varepsilon_{in}(r,q) + \varepsilon_{ex}(r,q),
\label{eq:(2-18)}
\en
where $\varepsilon(q)$ is the total flat-space energy of the ESS solution. The useful formula

\be
\varepsilon(q) = \dfrac{16\pi}{3} q^{3/2} \int_{0}^{E(r=0)} \left[y \cdot \frac{\partial \varphi}{\partial X}(X=y^{2})\right]^{-1/2} dy,
\label{eq:(2-18)ter}
\en
allows the explicit calculation of this total energy, once the expression of the lagrangian density is known \cite{dr08}. Using Eq.(\ref{eq:(2-12)}) we obtain the scaling law

\be
\varepsilon(q) = q^{3/2} \varepsilon(q=1),
\label{eq:(2-19)}
\en
where $\varepsilon(q=1)$ is the flat-space energy of the unit charge solution and is a universal constant for a given model. In terms of the interior integral of energy, equation (\ref{eq:(2-9)}) can be alternatively rewritten as

\be
\lambda(r,q,M) = 1 - \frac{2(M - \varepsilon(q))}{r} - \frac{2\varepsilon_{in}(r,q)}{r}.
\label{eq:(2-18)bis}
\en
The presence of this finite energy, as a parameter to which the ADM mass can be compared, leads to the existence of special solutions for which the total energy of the configuration, as defined by the ADM mass, equals the electrostatic energy in absence of gravitation. We shall call these solutions ``critical".

The function $\varepsilon_{ex}(r,q)$, for fixed $q$, is again monotonically decreasing and concave, whereas $\varepsilon_{in}(r,q)$ is monotonically increasing and convex. Returning to Eq.(\ref{eq:(2-9)}) we see that the roots of $\lambda(r,q,M) = 0$, giving the radii of the horizons for fixed $q$ and $M$, can be obtained (when non-vanishing) by solving the same equation (\ref{eq:(2-13)}). But now the function $\varepsilon_{ex}(r,q)$ reaches the finite value $\varepsilon(q)$ at $r=0$. Let us analyze separately the cases A1 and A2.

\underline{\textbf{Case A1}}: Near the origin the field behaves as $E(r,q) \sim \eta(q) r^{p}$, where $-1 < p < 0$ and $\eta(q) = \left(\frac{\beta}{q}\right)^{p/2}$ (see Eq.(\ref{eq:(2-7)})). From Eq.(\ref{eq:(2-11)}) we obtain for $\varepsilon_{ex}(r,q)$ the behaviour

\be
\varepsilon_{ex}(r,q) \sim \varepsilon(q) - \frac{16\pi q \eta(q)}{(2-p)(1+p)} r^{p+1},
\label{eq:(2-20)}
\en
around the center, where its slope diverges. The study of the intersection points of the curve $y = \varepsilon_{ex}(r,q)$ (for fixed $q$) with the beam of straight lines $y = M - r/2$ gives the masses and radii of the horizons and leads to the following classification of the solutions (see Fig.3):

\textbf{(A1-1)} There is again a family of extreme black hole solutions (with parameter $q$) whose radii $r_{hextr}(q)$ and ADM masses $M_{extr}(q)$ are given by the same equations (\ref{eq:(2-14)}) and (\ref{eq:(2-14)bis}) as in the divergent-energy case of section III. These are increasing functions of the charge, ranging from $0$ to $\infty$ as $q$ increases in the same interval, as can be shown from the scaling law (\ref{eq:(2-12)}) for the integral of energy.

\textbf{(A1-2)} Again, for configurations with values of $M < M_{extr}(q)$ there are no horizons and the solutions have a naked singularity at the center.

\textbf{(A1-3)} For $M_{extr}(q) < M < \varepsilon(q)$ the solutions exhibit two horizons. As $M - \varepsilon(q) \rightarrow 0^-$ the radii of the inner horizons of the sequence of solutions parameterized by $M$ goes to zero and the radii of the outer horizons approach a regular limit value $r_{hcrit}(q)$, given by the non-vanishing solution of the equation $r = 2\varepsilon_{in}(r,q)$. There is a singularity of metric function $\lambda$ at the center in this limit case, which steps from $+\infty$ to $-\infty$ on both sides of the vanishing-radius inner horizon. In both (A1-2) and (A1-3) cases the limit $M \rightarrow M_{extr}(q)$ is singular.

\textbf{(A1-4)} When $M = \varepsilon(q)$ we have (see Eqs.(\ref{eq:(2-18)bis}) and (\ref{eq:(2-20)}))

\be
\lambda(r,q) = 1 - \frac{2\varepsilon_{in}(r,q)}{r} \rightarrow -\infty \hspace{.5cm} as \hspace{.5cm}r \rightarrow 0,
\label{eq:(2-21)}
\en
and only the outer horizon of radius $r_{hcrit}(q)$ survives (see Fig.4). These solutions coincide, for $r > 0$, with the limit solutions of the sequence of the previous case (A1-3), but differ by the nature of the singularity at the center.

\textbf{(A1-5)} For $ M > \varepsilon(q)$ there is a unique horizon whose radius, obtained from Eq.(\ref{eq:(2-13)}), increases with $M$ from $r_{h}=r_{hcrit}(q)$ when $M=\varepsilon(q)$, becoming $r_{h} \sim 2M$ for large $M$. As $M - \varepsilon(q) \rightarrow 0^{+}$ (for fixed $q$) the sequence converges continuously to the critical solution of the case (A1-4) \textit{for any} $r$.

As an example of this family let us consider the effective action for QED of Euler and Heisenberg \cite{EH36}, whose lagrangian density takes the form

\be
\varphi(X,Y) = \frac{X}{2} + \mu\left(X^{2} + \frac{7}{4}Y^{2}\right),
\label{eq:(2-22)}
\en
where $\mu$ is a positive constant. This model satisfies the admissibility conditions and is a particular case of a family of polynomial lagrangian densities, all belonging to this class A1, whose ESS solutions in flat space were extensively analyzed in Ref.\cite{dr08}. The ESS solutions in this case are obtained from Eq.(\ref{eq:(2-7)}), which now takes the form

\be
2\mu E^{3}(r,q) + \dfrac{1}{2} E(r,q) = \dfrac{q}{r^{2}},
\label{eq:(2-22)bis}
\en
and can be solved explicitly through the Tartaglia formula, leading to

\be
E(r,q) = \left[\dfrac{v}{r^{2}} + \sqrt{\Delta}\right]^{1/3} + \left[\dfrac{v}{r^{2}} - \sqrt{\Delta}\right]^{1/3},
\label{eq:(2-22)ter}
\en
where $\Delta = \dfrac{v^{2}}{r^{4}} + u^{3}>0$, $u = \frac{1}{12\mu}$ and $v = \frac{q}{4\mu}$. Near the center these fields behave as $E(r \rightarrow 0,q) \sim (\frac{q}{2\mu})^{1/3} r^{-2/3}$ (case A1) and are asymptotically coulombian: $E(r \rightarrow \infty,q) \sim \frac{2q}{r^{2}}$ (case B2). These behaviours can be also immediately deduced by the simple inspection of the lagrangian density (\ref{eq:(2-22)}) and Eq.(\ref{eq:(2-7)}). The central field behaviour ($p=-2/3$), together with the admissibility conditions endorse the decreasing and concave character of the exterior integral of energy. Consequently, the statements A1-1 to A1-5 must hold for the gravitating version of this model. Let us calculate some characteristic magnitudes in order to obtain the quantitative description of the G-ESS configurations.

The energy density obtained from Eqs.(\ref{eq:(2-6)}) and (\ref{eq:(2-22)}) can be written, in terms of the field, as

\be
T_{0}^{0}(r,q) = E^{2}(r,q)\left[3\mu E^{2}(r,q) + \dfrac{1}{2}\right],
\label{eq:(2-22)s}
\en
and its central and asymptotic behaviours are

\be
T_{0}^{0}(r \rightarrow 0,q) \sim \dfrac{3 q^{4/3}}{(16\mu)^{1/3}} r^{-8/3} \hspace{.1cm};\hspace{.1cm}
T_{0}^{0}(r \rightarrow \infty,q) \sim \dfrac{2q^{2}}{r^{4}},
\label{eq:(2-22)ss}
\en
confirming that the slope of the integral of energy $\varepsilon_{ex}(r,q)$ diverges as $r \rightarrow 0$ (see Fig 3).

Using Eq.(\ref{eq:(2-18)ter}) the total flat-space energy associated to these solutions takes the form

\be
\varepsilon(q) = \frac{16\pi q^{3/2}}{3} \int_{0}^{\infty} \frac{dy}{\sqrt{y(1/2 + 2\mu y^{2})}} = \frac{8\pi q^{3/2}}{3\mu^{1/4}} B(\frac{1}{4},\frac{1}{4}),
\label{eq:(2-22)q}
\en
where $B(x,y) = \int_{0}^{1} t^{x-1}(1-t)^{y-1} dt$ is the Euler integral of first kind.
The extreme black hole solutions must satisfy Eq.(\ref{eq:(2-14)}), which now reads

\be
3\mu E^{4} + \dfrac{E^{2}}{2} = \dfrac{q}{8\pi r^{2}},
\label{eq:(2-22)se}
\en
leading to the relation

\be
E_{extr}^{2}(q) = \dfrac{1}{12\mu}\left(\sqrt{1 + \frac{6\mu}{\pi r^{2}_{hextr}(q)}} - 1\right)
\label{eq:(2-22)e}
\en
between the radii of the horizons of the extreme black holes and the values of the electrostatic field there. Let us introduce, by convenience, the auxiliary variable

\be
\chi = \sqrt{1 + \frac{6\mu}{\pi r^{2}_{hextr}(q)}} - 1,
\label{eq:(2-22)t}
\en
which is a monotonically decreasing and concave function of $r_{hextr}$, ranging from $\infty$ to $0$ as $r_{hextr}$ ranges from $0$ to $\infty$. Another independent relation between $E_{extr}(q)$ and $\chi$ is given by the first integral (\ref{eq:(2-22)bis}), and the elimination of $E_{extr}(q)$ between both equations leads to

\be
\dfrac{12 \pi^{2} q^{2}}{\mu} \chi = \left(1 + \dfrac{1}{\chi + 2}\right)^{2}.
\label{eq:(2-22)n}
\en
By drawing the functions of $\chi$ defined by the both sides of this equation (a beam of straight lines with slope proportional to $q^{2}$, cutting once a squared hyperbola) it is obvious that there is a unique solution $\chi(q) > 0$, which is a decreasing function of $q$, ranging from $\infty$ to $0$ as $q$ ranges from $0$ to $\infty$. Hence $r_{hextr}(q)$ is an increasing function of $q$ in the same interval, in agreement with the statement A1-1. Moreover, Eqs.(\ref{eq:(2-22)n}) and (\ref{eq:(2-22)t}) define explicitly the function $q(r_{hextr})$, giving the \textit{exact} relation between the charge and the horizon radius of the extreme black holes. The associated mass $M_{extr}(q)$ can now be obtained from Eq.(\ref{eq:(2-14)bis}) (after integration of the field  (\ref{eq:(2-22)ter}), leading to the form of the potential $A_{0}(r,q)$) and, as easily seen, it is also an increasing function of $q$ or $r_{hextr}(q)$, ranging from $0$ to $\infty$. It is now easy to verify that solutions with $M < M_{extr}(q)$, $M_{extr}(q) < M < \varepsilon(q)$ and $M > \varepsilon(q)$ lead to naked singularities, two-horizons black holes and single-horizon black holes, respectively, the previous equations allowing the quantitative determination of the parameters of these configurations.

\begin{figure}[h]
\begin{center}
\includegraphics[width=8.6cm,height=6.2cm]{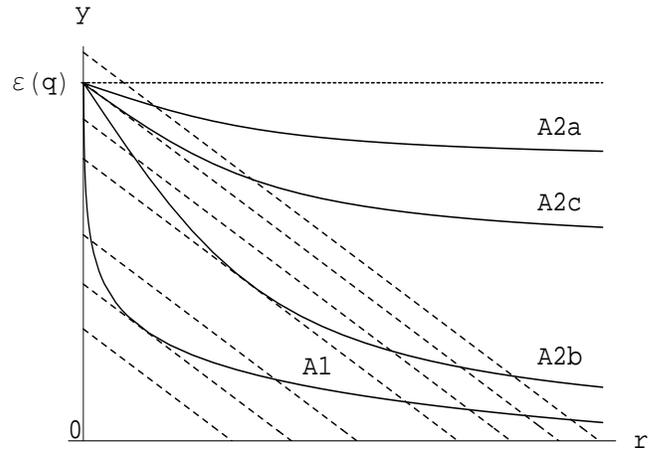}
\caption{\label{fig:epsart} Exterior integral of energy $\varepsilon_{ex}(r,q)$ for the two different cases of NEDs with (flat-space) finite-energy ESS solutions. The bottom curve corresponds to models with ESS fields divergent (but integrable) at the center (case A1). In this case the slope of the curve at the origin diverges. The remaining curves are representative of ESS solutions of NEDs leading to a finite value of the fields at the center (A2 cases). They correspond to the slopes at the center associated to different values of the electric charge ($\varepsilon_{ex}'(r=0)=-8 \pi aq \gtreqless 1/2$, see Eq.(\ref{eq:(2-23)})). The cut points with the straight (dashed) lines $y=(M-r/2)$ give the horizons of the different black hole solutions (see the discussion in the text). For the sake of simplicity curves corresponding to different values of the charge $q$ have been superimposed on the same figure, in such a way that all of them begin at the same point $\varepsilon(q)$. Consequently, each curve should be viewed as if it were represented in a different figure. Note that the slope of the beam of straight lines is not affected by this trick.}
\end{center}
\end{figure}

\underline{\textbf{Case A2}}: As above mentioned, this case corresponds to $p = 0$ near the origin and the ESS field solutions behave there as in Eq.(\ref{eq:(2-7)ter}). Before using Eq.(\ref{eq:(2-13)}) in order to analyze the structure of the G-ESS solutions we must study the behaviour of $\varepsilon_{ex}(r,q)$ around $r=0$. Using Eqs.(\ref{eq:(2-11)}), (\ref{eq:(2-17)}) and (\ref{eq:(2-7)ter}) we obtain the expression

\be
\frac{d \varepsilon_{ex}}{dr}\Big\vert_{r=0} = - 4\pi r^{2}T_{0}^{0}\Big\vert_{r=0} = -8\pi aq,
\label{eq:(2-23)}
\en
for the slope of $\varepsilon_{ex}(r,q)$ at the center. Moreover, the behaviour of the gravitational potential around the center can be obtained from Eqs.(\ref{eq:(2-18)bis}), (\ref{eq:(2-11)}) and (\ref{eq:(2-7)ter}) and reads

\be
\lambda(r) \sim 1 - 16\pi aq - \frac{2(M - \varepsilon(q))}{r} + \frac{32\pi bq}{(\sigma+1)(2-\sigma)}r^{\sigma}
+ \Delta r^{2},
\label{eq:(2-24)}
\en
for $\sigma \neq 2$ and

\bea
\lambda(r) &\sim& 1 - 16\pi aq -
\frac{2(M - \varepsilon(q))}{r}+ \nonumber \\&+&\frac{8\pi bq}{3}
r^{2}\left(1 - 2 \ln(r)\right) + \Delta r^{2},
\label{eq:(2-24)bis}
\ena
for $\sigma = 2$, $\Delta $ being an integration constant. When $\sigma > 2$ this constant is given by $\Delta = \frac{8\pi q \varphi(a^{2},0)}{3}$ (see Eq.(\ref{eq:(2-7)six})). Otherwise the value of $\Delta$ is not relevant because the associated term is not dominant. Consequently, we must now analyze several possibilities
corresponding to solutions with different values of $q$, which we shall denote as A2a (corresponding to $q < 1/16\pi a$), A2b (for $q > 1/16\pi a$) and A2c (for $q = 1/16\pi a$). Let us discuss the behaviour of the configurations corresponding to different values of $M$ for each one of these cases.

$\bullet$ \textbf{(A2a)} If $16\pi aq < 1$ we have three classes of solutions (see Fig.3):

\textbf{(A2a-1)} $M < \varepsilon(q)$: The solutions have no horizons but only naked singularities at the center. As $M - \varepsilon(q) \rightarrow 0^{-}$ the solutions, parameterized by $M$, converge towards a limit field with a discontinuity of $\lambda(r)$ at the center, which steps from $+\infty$ to the finite value $1 - 16\pi aq > 0$. For $r > 0$, $\lambda(r)$ converges towards the form of the critical case (A2a-2) below.

\textbf{(A2a-2)} $M = \varepsilon(q)$: In this case $\lambda(0) = 1 - 16\pi aq > 0$ is finite and $\lambda(r)$ remains regular and positive everywhere. The behaviour of the derivative of the metric near the center can be calculated from the expression (\ref{eq:(2-24)}) and reads (for $\sigma \neq 2$)

\be
\frac{d \lambda}{dr} \sim \frac{32\pi \sigma bq}{(\sigma+1)(2-\sigma)}r^{\sigma-1} + 2r\Delta,
\label{eq:(2-25)}
\en
which diverges for $\sigma < 1$, vanishes for $\sigma > 1$ and takes a positive finite value for $\sigma = 1$ (see Fig.4). Equation (\ref{eq:(2-24)bis}) shows that this derivative also vanishes for $\sigma = 2$. These are naked singularities whose structure at $r=0$ differs from that of the limit field defined in (A2a-1), but both fields coincide for $r>0$ (see Fig.4).

\textbf{(A2a-3)} $M > \varepsilon(q)$: The solutions have a unique and non-degenerate horizon. The radius of the horizon $r_{h}$ is related to the ADM mass through Eq.(\ref{eq:(2-13)}). At the center $\lambda(0) = -\infty$. As $M - \varepsilon(q) \rightarrow 0^{+}$, $r_{h} \rightarrow 0$ and the solutions, parameterized by $M$, converge towards a kind of ``\textit{black point}", with a discontinuity of $\lambda(r)$ at the center, which steps from $-\infty$ to zero on the internal side of the horizon ($r_{h}^{<}$), and from zero to $1 - 16\pi aq$ on the external side ($r_{h}^{>}$). For $r>0$ this limit solution is regular and coincides with the one of the critical case (A2a-2).

$\bullet$ \textbf{(A2b)} When $16\pi aq > 1$ we have five classes of solutions (see Fig.3 and the small frame of Fig.4):

\textbf{(A2b-1)} Now, as in the case A1, there is a family of extreme black hole solutions (with parameter $q$) whose radii $r_{hextr}(q)$ and ADM masses $M_{extr}(q)$ are the solutions of Eqs.(\ref{eq:(2-14)}) and (\ref{eq:(2-14)bis}) (see curve B in the small frame of Fig.4). The analysis of these equations, using the scaling law (\ref{eq:(2-12)}) for the integral of energy, shows that $r_{hextr}(q)$ and $M_{extr}(q)$ are monotonically increasing functions of $q$, ranging from $0$ to $\infty$ and from $\varepsilon(\frac{1}{16\pi a})$ to $\infty$, respectively, as $q$ ranges from $q=\frac{1}{16\pi a}$ to $\infty$.

\textbf{(A2b-2)} For $M < M_{extr}(q)$: There are no horizons and the solutions have a naked singularity at the center, analogue to the one of case (A1-2). As in the cases of RN and (A1-2), the limit $M \rightarrow  M_{extr}(q)$ is singular (see curve A in Fig.4).

\textbf{(A2b-3)} For $M_{extr}(q) < M < \varepsilon(q)$: As in case (A1-3) the solutions have two horizons. As $M - \varepsilon(q) \rightarrow 0^{-}$ the radii of the inner horizons vanish and the sequence of solutions converges to black holes with an external event horizon and an infinite jump of the metric at the center, which passes from $+\infty$ to the finite value $\lambda(0^{+}) = 1 - 16\pi aq < 0$ on both sides of the (vanishing-radius) inner horizon (see curves C and D in Fig.4). As in the RN case, the limit $M - M_{extr}(q) \rightarrow 0^{+}$ is singular.

\textbf{(A2b-4)} When $M = \varepsilon(q)$: As in case (A2a-2) $\lambda(0) = 1 - 16\pi aq < 0$ is finite (but now negative), and $\lambda(r)$ is regular everywhere and vanishes at some $r=r_{hcrit}$, where there is a unique horizon. The derivative of the metric near the center takes the same form (\ref{eq:(2-25)}) as in the (A2a-2) case, with the same behaviour as a function of $\sigma$. These are now black holes which coincide, for $r>0$, with the limit cases $M - \varepsilon(q) \rightarrow 0^{-}$ of the sequence of solutions (A2b-3) (it coincides also with the limit of the sequence (A2b-5) below, as $M - \varepsilon(q) \rightarrow 0^{+}$) but the
structure of the singularities at the center is different in each case (see curves D and E in Fig.4).

\textbf{(A2b-5)} For $M > \varepsilon(q)$: There is now a unique horizon with a radius ranging from $r_{h}=r_{hcrit}$ to $\infty$ as $M$ increases. The sequence of solutions converges as $M - \varepsilon(q) \rightarrow 0^{+}$, for $r>0$, to the critical case (A2b-4), but now $\lambda(r)$ steps from $-\infty$ to the negative value $\lambda(0) = 1 - 16\pi aq < 0$ at the center (see curves E and F in the small frame of Fig.4).

$\bullet$ \textbf{(A2c)} When $16\pi aq = 1$ the charge is fixed, Eq.(\ref{eq:(2-14)}) is satisfied at $r = 0$ and  Eqs.(\ref{eq:(2-24)}) and (\ref{eq:(2-24)bis}) get simplified. We can distinguish three cases (see Fig.3):

\textbf{(A2c-1)} For $M = \varepsilon(q)$: In this case $\lambda(r)$ is regular everywhere and vanishes at $r=0$, where the condition for the radius of extreme black holes (\ref{eq:(2-14)}) holds. There are three behaviours for the slope of $\lambda(r)$ at $r=0$ (depending on the values of $\sigma \lesseqqgtr 1$) given by Eq.(\ref{eq:(2-25)}). These critical solutions are \textit{extreme black points} (see Fig.4).

\textbf{(A2c-2)} For $M < \varepsilon(q)$: We have naked singularities with the same behaviour as in case (A2a-1) but in the limit $M - \varepsilon(q) \rightarrow 0^{-}$ an extreme black point appears with a discontinuity of $\lambda(r)$, which steps from $+\infty$ to zero at the center. This limit is singular.

\textbf{(A2c-3)} For $M > \varepsilon(q)$: The behaviour is similar as in (A2a-3) and we have a unique horizon. At the center $\lambda(0) = -\infty$, and the solutions converge to black points as $M - \varepsilon(q) \rightarrow 0^{+}$. Once more the limit of the sequences coincides with the solutions of the critical case (A2c-2) for $r>0$, but they have point-like discontinuities at the center.

\begin{figure}[h]
\begin{center}
\includegraphics[width=8.6cm,height=6.6cm]{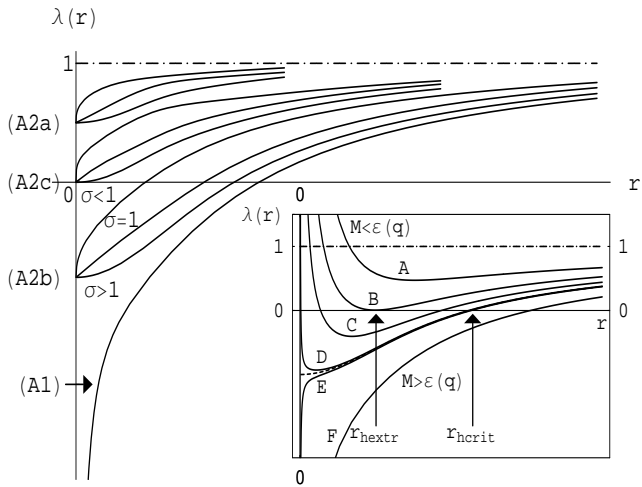}
\caption{\label{fig:epsart} Qualitative behaviour of the metric function $\lambda(r)$ for the critical configurations ($M = \varepsilon(q)$) in the cases A1 ($\lambda(0) \rightarrow -\infty,$ case A1-4 in the text), A2a ($\lambda(0) = 1 - 16\pi a q > 0,$ case A2a-2 in the text), A2b ($\lambda(0) = 1 - 16\pi a q < 0,$ case A2b-4 in the text) and A2c ($\lambda(0) = 1 - 16\pi a q = 0,$ case A2c-1 in the text). The three curves associated to each (A2) case correspond to the three ranges of the parameter $\sigma \lesseqqgtr 1,$ which determine their slopes at $r=0$ (see Eq.(\ref{eq:(2-25)})). The small frame displays the metric function for different values of the ADM mass in the case A2b with $\sigma > 1$. The dashed line in this frame corresponds to the critical configuration ($M = \varepsilon(q),$ case A2b-4 in the text). The upper curves ($M < \varepsilon(q)$) correspond to naked singularities (A), extreme black holes (B) and two-horizon black holes (C and D). The lower curves (E and F, with $M > \varepsilon(q)$) correspond to the single-horizon black holes. Curves D and E show the approximation to the critical configuration (for $r>0$) in the limits $M - \varepsilon(q) \rightarrow 0^{\mp}$, respectively, illustrating the formation of the isolated discontinuities of the limit configurations at $r=0$. The horizon radii for the extreme ($r_{hextr}$) and critical ($r_{hcrit}$) black holes are indicated. Similar figures to the one of the small frame can be straightforwardly displayed for the other values of $\sigma$ and for the cases A1, A2a and A2c, illustrating the discussion of section IV.}
\end{center}
\end{figure}

As a well known example of this family let us consider the case of the original Born-Infeld model (whose gravitating ESS solutions were studied in Ref.\cite{garcia84}), defined by the lagrangian density

\be
L_{BI} = \varphi_{BI}(X,Y) = \frac{1 - \sqrt{1 - \mu^{2}X - \frac{\mu^{4}}{4}Y^{2}}}{\mu^{2}/2},
\label{eq:(2-26)}
\en
where $\mu$ is a constant, assumed positive. As $\mu \rightarrow 0$ this function reduces to the Maxwell lagrangian density $\varphi_{M}(X) = X$. It is straightforward to check that this model satisfies the admissibility conditions of section II. The ESS solutions take the form

\be
E(r,q) = \dfrac{q}{\sqrt{r^{4} + \mu^{2} q^{2}}},
\label{eq:(2-27)}
\en
which are asymptotically coulombian, reduce to the Coulomb field as $\mu \rightarrow 0$ and behave around the center as $E(r,q) \sim \frac{1}{\mu}-\frac{r^4}{2\mu^3 q^2} $. Thus the parameters of the expansion (\ref{eq:(2-7)ter}) are $a = 1/\mu$ (the maximum strength of the field), $\sigma = 4$ and $b = (2\mu^{3} q^{2})^{-1}$ (which satisfies the relation (\ref{eq:(2-7)quart})). The energy density for this solution reads

\be
T_{0}^{0}(r,q) = 2\dfrac{\sqrt{r^{4} + \mu^{2} q^{2}} - r^{2}}{\mu^{2} r^{2}},
\label{eq:(2-28)}
\en
and the exterior integral of energy is given by

\bea
\varepsilon_{ex}(r,q) &=& \frac{8\pi r}{3\mu^{2}} \Big[r^{2} - \sqrt{r^{4} + \mu^{2} q^{2}} + \nonumber \\ &+& \frac{2\mu^{2}q^{2}}{r^{2}} F_{1}\left(\frac{1}{4},\frac{1}{2},\frac{5}{4},\frac{-\mu^{2}q^{2}}{r^{4}}\right)\Big],
\label{eq:(2-29)}
\ena
where $F_{1}(x,y,z,u)$ is the Gaussian hypergeometric function. As a consequence of the admissibility conditions, $\varepsilon_{ex}(r,q)$ is a monotonically decreasing and concave function for fixed $q$, which can be directly verified from its explicit expression. The total flat-space energy of the field takes the form

\be
\varepsilon(q) = \frac{8 \pi^{5/2} q^{3/2}}{3\mu^{1/2} \Gamma\left(\frac{3}{4}\right)^{2}},
\label{eq:(2-30)}
\en
where $\Gamma(x)$ is the Euler integral of second kind.

If we solve now the equation (\ref{eq:(2-14)}) for the extreme black holes we are lead to

\be
r_{hextr}(q) = \sqrt{\dfrac{(16\pi q - \mu)(16\pi q +
\mu)}{32\pi}},
\label{eq:(2-31)}
\en
which requires the condition $16\pi q \geq \mu$ to be fulfilled, in agreement with the preceding analysis (solutions A2b and A2c). The masses of these extreme black holes can be now immediately found by using Eqs.(\ref{eq:(2-13)}), (\ref{eq:(2-29)}) and (\ref{eq:(2-31)}). As stated for general A2b solutions, the radii and masses of these black holes are increasing functions of the charge. The radii range from $0$ to $\infty$ as $q$ ranges from $\frac{\mu}{16\pi}$ to $\infty$, whereas the masses range from $\varepsilon(\frac{\mu}{16\pi})$ (extreme black point solution) to $\infty$ in the same range of $q$.

The verification of the remaining general statements for this BI example is now straightforward.

\section{Conclusions and perspectives}

Let us summarize the main results found until now. We have considered the full family of NEDs whose lagrangian densities are functions of the two gauge invariants, restricted by some admissibility conditions and exhibiting asymptotically vanishing ESS solutions. When minimally coupled to gravity, we have established that the central field behaviour of these solutions (or, equivalently, the behaviour of the lagrangian densities on the boundary of their domain of definition at $Y=0$) fully characterizes the structure of their gravitating versions. In the case of NEDs with energy-divergent ESS solutions in flat space, the behaviour of the G-ESS solutions is qualitatively the same as the one of the RN solutions of Einstein-Maxwell field equations. In the case of NEDs with (flat-space) finite-energy ESS solutions, qualitatively different features appear. Aside from naked singularities, extreme black holes with a single horizon and two-horizons black hole solutions (as in the RN case), there are now non-extreme single-horizon black hole solutions and black-point solutions. In all cases there are, \textit{at most}, two horizons. The presence of the ``critical" solutions must be stressed. They can lead to naked singularities, as in the case (A2a-2), single-horizon black holes, as in cases (A1-4) and (A2b-4) or extreme black points as in the case (A2c-1). In all cases, the sequences of solutions (depending on the two integration constants $M$ and $q$) converge, as $M \rightarrow \varepsilon(q)$, at fixed $q$, to limit metrics which coincide with the critical solutions for $r>0$. Nevertheless, their structures at the center differ by singular ($\delta$-like or step) terms. This behaviour, which is similar to the one encountered in the Schwarzschild and RN solutions \cite{ortin}, deserves a more careful analysis in order to elucidate the nature of the eventual point-like sources of the diverse solutions.

At the center of these solutions there is always a curvature singularity, as established by the Bronnikov theorem \cite{bronnikov80,bronnikov0,bronnikov01}. In fact this theorem applies only for asymptotically coulombian fields (case B2). Nevertheless it can be easily generalized to the case B1. The argumentation of the theorem, as presented in Ref.\cite{bronnikov0}, fails in the case B3, but the regularity of the manifold at the center of the solutions can be excluded also in this case, for admissible models, by an alternative argument involving the positivity of the energy (see Ref.\cite{dr09}).

Asymptotically, if the fields behave as $E(r) \sim r^{p}$ (with $p < -1$, in order to endorse the convergence of the integral of energy $\varepsilon_{ex}(r,q)$), the corresponding metrics behave as

\be
\lambda(r) \sim 1- \frac{2M}{r} + \frac{32\pi q }{(p-2)(p+1)} r^{p} + \cdot
\label{eq:(2-16)}
\en
and approach asymptotic flatness as the Schwarzschild metric, excepting for solutions with vanishing ADM mass (which are always naked singularities) for which the gravitational field ($1 - \lambda(r)$) vanishes asymptotically at the same cadence as $E(r)$. Obviously, solutions with $p = -2$ (case B2) exhibit the same asymptotic behaviour as the RN one.

Beyond the RN solution and the gravitating EH and BI models, employed here as representative of the families of admissible theories in this context, the elaboration of other explicit examples reduces to the finding of the lagrangian densities $\varphi(X,Y)$ satisfying the conditions displayed in Fig.1. This method, developed in Ref.\cite{dr08}, exhausts the class of this kind of admissible models.

The thermodynamic analysis of Schwarzschild and RN black holes can be generalized to the present models. The results of Ref.\cite{rasheed97} show how the laws of thermodynamics can be consistently formulated for non-linear electrodynamic black holes. A discussion of the thermodynamic features of the gravitating NED models considered here is beyond the scope of this paper. This issue will be tackled elsewhere \cite{dr09}.

Finally, some comments on the stability of the G-ESS solutions are in order. The necessary and sufficient condition for the linear stability of the ESS solutions of NEDs in flat space has been obtained in Ref.\cite{dr08} and requires the inequality

\be
\frac{\partial \varphi}{\partial X} \geq 2X\frac{\partial^{2} \varphi}{\partial Y^{2}},
\label{eq:(2-32)}
\en
to be satisfied by the lagrangian densities in the domain of definition of the solution. It is not clear a priory whether the effect of the gravitational field improves the stability of solutions which are unstable in flat space, or whether new conditions should be added to the lagrangian densities of the admissible NEDs supporting stable ESS solutions in flat space, to maintain stability of their gravitating generalizations. The analysis of this issue is in progress \cite{dmr}. \\

\begin{center}
\textbf{ACKNOWLEDGMENTS} \\
\end{center}

We are indebted to Drs. S. Hacyan, J. M. Martin-Garc\'ia and A. R. Plastino for useful discussions. D. R. G was supported by a FICYT (Spain) grant, project number IB08-154.


\begin{thebibliography}{99}

\bibitem{BI34} M. Born and L. Infeld, Proc. R. Soc. London. A \textbf{144}, 425
(1934).

\bibitem{fradkin85} E. S. Fradkin and A. A. Tseytlin, Phys. Lett. B \textbf{163}, 123 (1985); A. Abouelsaood, C. G. Callan, Jr., C. R. Nappi and S. A. Yost, Nucl.
Phys. B \textbf{280}, 599 (1987);  R. G. Leigh, Mod.
Phys. Lett. A \textbf{4}, 2767 (1989); A. A. Tseytlin, Nucl. Phys. B \textbf{501}, 41 (1997); D. Brecher, Phys. Lett. B \textbf{442}, 117 (1998).

\bibitem{deser76} G. H. Derrick, J. Math. Phys. \textbf{5}, 1252 (1964); S. Deser, Phys. Lett. B \textbf{64}, 463 (1976);
S. Coleman, Commun. Math. Phys. \textbf{55}, 113 (1977).

\bibitem{bart88} R. Bartnik and J. McKinnon, Phys. Rev. Lett. \textbf{61}, 141 (1988).

\bibitem{v-d-g} M. S. Volkov and D. V. Gal'tsov, Phys. Rept. \textbf{319}, 1 (1999).

\bibitem{lee87} T. D. Lee, Phys. Rev. D \textbf{35}, 3637 (1987); T. D. Lee and Y. Pang, Phys. Rept. \textbf{221}, 252 (1992).

\bibitem{hoffmann37} B. Hoffmann, Phys. Rev. \textbf{47}, 877 (1935); B. Hoffmann and L. Infeld, Phys. Rev. \textbf{51}, 765 (1937).

\bibitem{garcia84} A. Garcia, H. Salazar and J. F. Plebanski, Nuovo. Cim. \textbf{84}, 65 (1984); M. Demianski, Found. Phys. \textbf{16}, 187 (1986); D. L. Wiltshire, Phys. Rev. D \textbf{38}, 2445 (1988).

\bibitem{oliveira94} H. P. de Oliveira, Class. Quant. Grav.
\textbf{11}, 1469 (1994).

\bibitem{tamaki00} T. Tamaki and T. Torii, Phys. Rev. D \textbf{62}, 061501(R) (2000); G. Clement and D. V. Gal'tsov, Phys. Rev. D \textbf{62}, 124013 (2000).

\bibitem{fernando03} S. Fernando and D. Krug, Gen. Rel. Grav. \textbf{35}, 129 (2003); S. Fernando, Phys. Rev. D \textbf{74}, 104032 (2006).

\bibitem{dey04} T. K. Dey, Phys. Lett. B \textbf{595}, 484 (2004);
R. -G. Cai, D. -W. Pang and A. Wang, Phys. Rev. D \textbf{70}, 124034 (2004).

\bibitem{yajima01} H. Yajima and T. Tamaki, Phys. Rev. D \textbf{63}, 064007 (2001).

\bibitem{soleng95} H. H. Soleng, Phys. Rev. D \textbf{52}, 6178 (1995).

\bibitem{hassaine07} M. Hassaine and C. Martinez, Phys. Rev. D \textbf{75}, 027502 (2007);
M. Hassaine and C. Martinez, Class. Quant. Grav. \textbf{25},
195023 (2008).

\bibitem{bronnikov80} K. A. Bronnikov, V. N. Melnikov, G. N. Shikin and K. P. Staniukowicz, Ann. Phys. \textbf{118}, 84 (1979).

\bibitem{bronnikov01} K. A. Bronnikov, Phys. Rev. D \textbf{63}, 044005 (2001).

\bibitem{burinskii} A. Burinskii and S. R. Hildebrandt, Phys. Rev. D \textbf{65}, 104017 (2002); I. Dymnikova, Class. Quant. Grav. \textbf{21}, 4417 (2004).

\bibitem{a-b} E. Ay\'on-Beato and A. Garc\'ia,
Phys. Rev. Lett. \textbf{80}, 5056 (1998).

\bibitem{bronnikov0} K. A. Bronnikov, Phys. Rev. Lett. \textbf{85}, 4641 (2000).

\bibitem{dr08} J. Diaz-Alonso and D. Rubiera-Garcia, Ann. Phys. \textbf{324}, 827 (2009).

\bibitem{carroll09} S. M. Carroll, M. C. Johnson and L. Randall, arXiv:0901.0931[hep-th].

\bibitem{ortin} T. Ortin, \emph{Gravity and strings} (Cambridge Monographs on Mathematical Physics, C.U.P., 2004).

\bibitem{EH36} W. Heisenberg and H. Euler, Z. Phys. \textbf{98}, 714 (1936); J. Schwinger, Phys. Rev. \textbf{82}, 664 (1951).

\bibitem{dr09} J. Diaz-Alonso and D. Rubiera-Garcia, work in progress.

\bibitem{rasheed97} D. A. Rasheed, arXiv:hep-th/9702087.

\bibitem{dmr} J. Diaz-Alonso, J. M. Martin-Garcia and D. Rubiera-Garcia, in preparation.

\end{thebibliography}
\end{document}